\documentclass{jpp}
\usepackage{graphicx}

\usepackage[utf8]{inputenc}
\usepackage[T1]{fontenc}
\usepackage{amsmath}

\shorttitle{Self-Consistent Full-Wave Fokker-Planck LHCD Simulation}
\shortauthor{S.J. Frank, J.P Lee, J.C. Wright, I.H. Hutchinson, and P.T. Bonoli}

\title{Verifying raytracing/Fokker-Planck lower-hybrid current drive predictions with self-consistent full-wave/Fokker-Planck simulations}

\author{S. J. Frank\aff{1}
  \corresp{\email{frank@psfc.mit.edu}},
  J. P. Lee\aff{2}
  J. C. Wright\aff{1},
  I. H. Hutchinson\aff{1},
  \and P.T. Bonoli\aff{1}
 }

\affiliation{\aff{1}Plasma Science and Fusion Center, Massachusetts Institute of Technology, 
Cambridge MA 02139, USA \aff{2}Department of Nuclear Engineering, Hanyang University, Seoul, South Korea}

\begin{document}

\maketitle
\begin{abstract}
Raytracing/Fokker-Planck (FP) simulations used to model lower-hybrid current drive (LHCD) often fail to reproduce experimental results, particularly when LHCD is weakly damped. A proposed reason for this discrepancy is the lack of "full-wave" effects, such as diffraction and interference, in raytracing simulations and the breakdown of raytracing approximation. Previous studies of LHCD using non-Maxwellian full-wave/FP simulations have been performed, but these simulations were not self-consistent and enforced power conservation between the FP and full-wave code using a numerical rescaling factor. Here we have created a fully-self consistent full-wave/FP model for LHCD that is automatically power conserving. This was accomplished by coupling an overhauled version of the non-Maxwellian TORLH full-wave solver and the CQL3D FP code using the Integrated Plasma Simulator. We performed converged full-wave/FP simulations of Alcator C-Mod discharges and compared them to raytracing. We found that excellent agreement in the power deposition profiles from raytracing and TORLH could be obtained, however, TORLH had somewhat lower current drive efficiency and broader power deposition profiles in some cases. This discrepancy appears to be a result of numerical limitations present in the TORLH model and a small amount of diffractional broadening of the TORLH wave spectrum. Our results suggest full-wave simulation of LHCD is likely not necessary as diffraction and interference represented only a small correction that could not account for the differences between simulations and experiment.
\end{abstract}

\section{Introduction}

Lower-hybrid current drive (LHCD) is a radiofrequency (RF) actuator used to efficiently drive current in tokamaks \citep{Fisch1987}. LHCD drives current via electron Landau damping of slow-waves on fast electrons with $v_{e} \ge 3 v_{the}$, in the lower-hybrid (LH) limit, $\Omega_i \ll \omega \ll \Omega_e$, where $\omega$ is the wave frequency, $\Omega_s = q_s B / m_s$ is the cyclotron frequency of species $s$, and $v_{the} = \sqrt{2T_e/m_e}$ is the electron thermal velocity \citep{Bonoli1985}. This drives current by direct parallel momentum injection and distortion of the perpendicular distribution such that the parallel resistivity is reduced in the direction of the plasma current \citep{Fisch1978,Fisch1980}. 

Despite a long history of experimental demonstrations \citep{Bernabei1982,Porkolab1984,Bartiromo1986,Moriyama1990,Jacquinot1991,Ide1992,Peysson2001,Wilson2009,Cesario2010,Wallace2011,Liu2015}, LHCD has remained difficult to predict \citep{Bonoli2014}. Difficulty predicting LHCD is attributed to the existence of a "spectral gap", whereby LH waves tend to damp despite being launched with a parallel phase velocity, $v_{ph\parallel} \sim 6-8 v_{the0}$, much larger than their linear damping condition of $v_{ph\parallel} \sim 3 v_{the0}$. A number of mechanisms that upshift the parallel refractive index $N_\parallel = k_\parallel c/ \omega = c/v_{ph,\parallel}$, where $k_\parallel$ is the parallel component of the wave-vector $\vec{k}$, closing the spectral gap have been proposed. These mechanisms include: toroidally induced increases in the parallel wave number and wave scattering from turbulence \citep{Bonoli1981,Bonoli1982,Decker2014,Biswas2020,Biswas2021}, parametric wave interactions \citep{Cesario2004,Decker2014}, and diffractional broadening of the wave spectrum \citep{Pereverzev1992,Wright2009,Wright2010,Shiraiwa2011,Wright2014}. Which $N_\parallel$ upshift mechanism is dominant in a given situation can profoundly affect wave damping and current drive efficiency. Therefore, it is important to understand if and when each of these different mechanisms is important. 

Lower-hybrid current drive is typically simulated using raytracing. Raytracing is derived by applying the WKB method, assuming that $kL \gg 1$, to the plasma wave equation yielding the ray equations. To calculate non-Maxwellian plasma response and current drive, raytracing simulations are coupled to a Fokker-Planck simulation by formulating a quasilinear diffusion coefficient using the method described in \citet{Bonoli1986}. However, raytracing neglects the effects of diffraction and the $kL \gg 1$ limit breaks down near cutoffs and ray caustics. Solving the wave equation directly using a "full-wave" simulation preserves the effects of diffraction and interference as well as accurately capturing the behavior of the wave at cutoffs and caustics, and comparison of raytracing to full-wave simulations can allow us to determine if raytracing is accurate despite not including these effects. In this work, we implemented the first fully self-consistent full-wave/FP model that allows us to calculate LHCD and compare to results obtained with a raytracing/FP model in Alcator C-Mod. Previous LHCD modeling work has included non-Maxwellian effects \citep{Wright2010,Wright2014,Shiraiwa2011,MeneghiniThesis}, however, these simulations were not performed fully self-consistently. They required adhoc numerical rescaling of the quasilinear diffusion coefficient to ensure the correct RF power was deposited in the FP calculation or employed an approximate method for calculating the non-Maxwellian dielectric tensor. We will describe the construction of our fully self-consistent non-Maxwellian full-wave/FP LHCD model and its application to modelling a set of Alcator C-Mod experiments.  

\section{The Non-Maxwellian TORLH/CQL3D Model}\label{sec:torlhcql3dcoupling}

In order to perform non-Maxwellian simulations of LHCD using a full-wave simulation code it is necessary to iterate calculations of wave propagation and damping obtained using the plasma Helmholtz equation in the lower-hybrid limit and a Fokker-Planck equation solver such as CQL3D \citep{Kerbel1985,Harvey1992}. This iteration solves the system:
\begin{eqnarray}
    \nabla \times \nabla \times \vec{E} &=& \epsilon_\perp \vec{E}_\perp + i\epsilon_{xy}(\hat{b}\times\vec{E}_\perp) + \epsilon_\parallel(f) E_\parallel \hat{b} \label{eq:helmholtz} \\
     \frac{Df}{Dt} &=& C(f) + Q(f,E_\parallel),\label{eq:fokkerplanck}
\end{eqnarray}
where $\vec{E}$ is the wave electric field, $f$ is the electron distribution function, $\hat{b} = \vec{B}_0/|B_0|$ is the background magnetic field vector, $C(f)$ is the collision operator, $Q(f,E_\parallel)$ is the divergence of the quasilinear flux, and $\underline{\underline{\epsilon}}(f)$ is the plasma dielectric tensor with components:
\begin{subequations}
\begin{align}
    \epsilon_\perp &\simeq S = \frac{1}{2}(R + L)\\
    \epsilon_{xy} &\simeq D = \frac{1}{2}(R - L)\\
    \epsilon_\parallel (f) &= 1 - \frac{\omega_{pe}^2}{\omega^2}\zeta_e^2\textrm{Re}\big\{\textrm{Z}^\prime(\zeta_e)\big\} - \frac{\omega_{pi}^2}{\omega^2} + \textrm{Im}[\chi_{zz} (f)],
\end{align}
\end{subequations}
where $S,D,R,L$ are the cold plasma dielectric components from \citet{Stix}, $\omega_{ps}=q_s^2 n_s / m_s \epsilon_0$ is the plasma frequency for species $s$, $\zeta_s = \omega/k_\parallel v_{ths}$, and $\textrm{Z}$ the plasma dispersion function \citep{FriedConte}. The non-Maxwellian imaginary correction to $\epsilon_\parallel (f)$ here is $\textrm{Im}[\chi_{zz} (f)]$, discussed further in Section \ref{sec:chizzlookup} and Appendix \ref{sec:chizzform}. TORLH calculates an E-field for a single toroidal mode $n_\phi$ by solving Helmholtz's equation (\ref{eq:helmholtz}) discretized with semi-spectral representation:
\begin{equation}\label{eq:torlhdisc}
    \vec{E} = e^{i(n_\phi\phi - \omega t)}\sum_{m=-\infty}^{+\infty} \vec{E}^{(m)}(\psi)e^{im\theta},
\end{equation}
 where, $n_\phi$ is the toroidal mode number, $m$ is the poloidal mode number, and $\vec{E}^{(m)}(\psi)$ is represented by a cubic Hermite interpolating polynomial finite element. The TORLH discretization is global along flux surfaces and precisely defines the parallel $k$ vector:
 \begin{equation}\label{eq:kpartorlh}
     k_\parallel = m (\hat{b}\cdot\nabla\theta) + n_\phi (\hat{b}\cdot\nabla\phi) = \frac{m}{n_\tau}\frac{B_\theta}{|B|} + \frac{n_\phi}{R}\frac{B_\phi}{|B|},
\end{equation}
 where $n_\tau$ is a metric coefficient that functions as a generalized minor radius. These properties of the TORLH discretization allow hot plasma effects, like Landau damping, to be resolved without approximations required by fully finite element discretizations \citep{Shiraiwa2011}. Our choice of discretization produces a block tridiagonal system of equations solved using a custom parallel block cyclic reduction solver \citep{Lee2014}. 

After the field solve, TORLH performs a post-processing step to obtain the quasilinear diffusion coefficient $D_{ql,\parallel}(E_\parallel)$, the component of the quasilinear term $Q(f,E_\parallel)$ dependent on the parallel electric field. The CQL3D quasilinear term may be written \citep{Petrov2016}:
\begin{equation}\label{eq:qlflux}
    Q(f,E_\parallel) = \left\{\frac{1}{u^2}\frac{\partial}{\partial u}\left(\textrm{B}_0\frac{\partial}{\partial u} + \textrm{C}_0\frac{\partial}{\partial \vartheta} \right) + \frac{1}{u^2 \sin\vartheta} \frac{\partial}{\partial \vartheta} \left(\textrm{E}_0\frac{\partial}{\partial u} + \textrm{F}_0\frac{\partial}{\partial \vartheta} \right) \right\}f.
\end{equation}
Here, $u$ is the momentum per rest mass, $p/\gamma m$, $\vartheta$ is the velocity-space pitch angle, and $\textrm{B}_0$, $\textrm{C}_0$, $\textrm{E}_0$, and $\textrm{F}_0$ are bounce averaged quasilinear diffusion coefficients. We may relate the \textit{local} $\textrm{B}$ to the $D_{ql,\parallel}$ from \citet{KennelEngelmann1966} as follows:
\begin{equation}\label{eq:D2B}
    \textrm{B} = u^2D_{uu} = u^2(\cos\vartheta)^2 D_{ql,\parallel},
\end{equation}
Then perform a normalized zero-orbit-width bounce average of $\textrm{B}$ to obtain $\textrm{B}_0$:
\begin{equation}\label{eq:bounceavg}
    \textrm{B}_0 = \lambda\langle \textrm{B} \rangle = v_{\parallel,0} \oint \frac{dl}{v_\parallel} \textrm{B} = v_{\parallel,0} \int_0^{2\pi} d\theta \mathcal{J} \frac{B(\theta)}{v_\parallel} \textrm{B},
\end{equation}
 \begin{figure}
  \centering
  \includegraphics[width=0.75\linewidth]{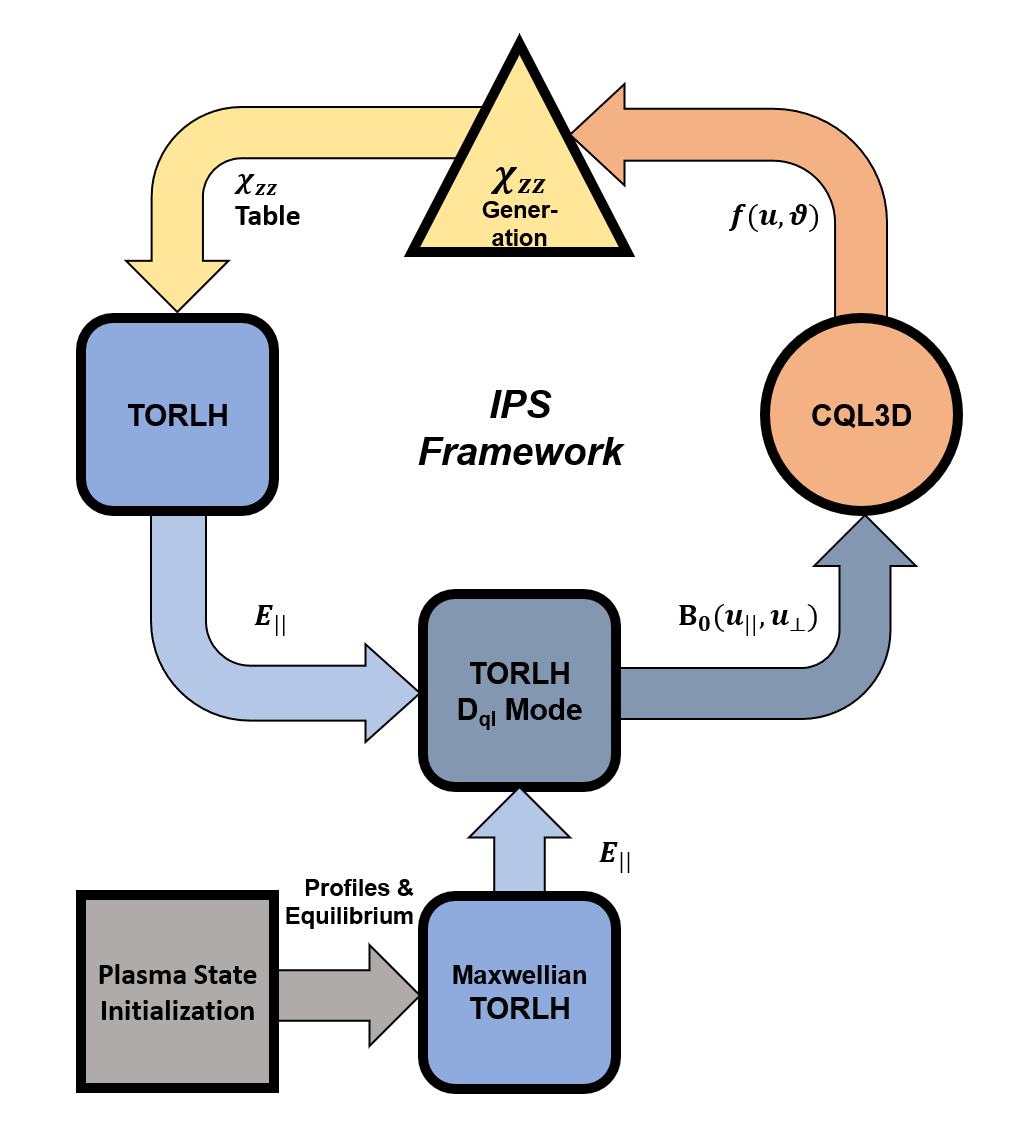}
  \caption{Flowchart showing an IPS driven iteration. The IPS creates a "plasma state" containing magnetic equilibrium and plasma profile information then initializes a Maxwellian TORLH simulation from that plasma state. TORLH and CQL3D are then iterated using the IPS framework supplemented by Python and FORTRAN wrappers that manage input files.}
\label{fig:ipsiter}
\end{figure}
where the integration element $dl$ is directed along the magnetic field line and $v_{\parallel,0}$ is the parallel velocity at the outboard midplane, and $\mathcal{J}$ is the Jacobian of the magnetic coordinate system. With $\textrm{B}_0$ we can obtain $\textrm{C}_0$, $\textrm{E}_0$, and $\textrm{F}_0$:

\begin{subeqnarray}
    \textrm{C}_0 = -\frac{\sin{\vartheta}}{u\cos{\vartheta}}\textrm{B}_0\\
    \textrm{E}_0 = -\frac{\sin^2\vartheta}{u\cos{\vartheta}}\textrm{B}_0\\
    \textrm{F}_0 = \frac{\sin^3\vartheta}{(u\cos{\vartheta})^2}\textrm{B}_0.
\end{subeqnarray}

Using the quasilinear diffusion computed by TORLH, CQL3D solves (\ref{eq:fokkerplanck}) to evolve the distribution for some amount of time. The evolved non-Maxwellian distribution function is passed back to a pre-processing routine that calculates a lookup table with values of the imaginary component of $\epsilon_\parallel$, $\textrm{Im}[\chi_{zz}]$ (the only component of the dielectric which meaningfully varies from the Maxwellian) and then TORLH is rerun using the updated non-Maxwellian dielectric. This process is repeated until power deposition profiles stop evolving and the driven current has risen to a steady state value indicating the system has converged. Convergence is typically obtained after $\sim 20-50$ iterations between TORLH and CQL3D. 

To facilitate simulation of multiple current drive scenarios using the TORLH/CQL3D iteration, shown in Figure~\ref{fig:ipsiter}, it was necessary to develop an automated framework for performing simulations. We created a set of Python and FORTRAN wrappers for the Integrated Plasma Simulator (IPS) \citep{Elwasif2010}. The IPS enabled the TORLH/CQL3D iterations to be performed at large scale with checkpointing and restart capabilities. Using the IPS, we successfully performed large scale TORLH/CQL3D simulations with $\sim 10,000$ cores on Cori at the National Energy Research Scientific Computing Center (NERSC) that could be iterated for up to 48 hours. However, initial simulations using the version of TORLH and the set-up from \citet{Wright2014} were not power conserving. In order to obtain physically and numerically self-consistent results, we needed to not only implement the improved TORLH boundary condition and field solver convergence criteria from \citet{Frank2022}, but we also had to implement significant improvements in the non-Maxwellian components of TORLH including: reformulation of the quasilinear diffusion coefficient, construction of a $\textrm{Im}[\chi_{zz}]$ lookup table, and the lookup table interpolation in TORLH. In the following sections we will describe the key modifications to the non-Maxwellian components of TORLH and develop convergence requirements for them.

\subsection{Quasilinear Diffusion Coefficient Formulation}\label{sec:dql}

The quasilinear diffusion coefficient, $D_{ql}$, formulation in TORLH was completely rewritten during the course of this work in order to implement a form that was power conserving. The TORLH diffusion coefficient is derived from the parallel component of the RF quasilinear diffusion coefficient in \citet{KennelEngelmann1966}: 
\begin{eqnarray} \label{eq:kedqlstrt}
    D_{ql\parallel} = \frac{\pi e^2}{2m_e^2}Re\Bigg\{\int dk_{\parallel1}\int dk_{\parallel2} \frac{v_\parallel^2}{c^2}\left(E_{\parallel 1} J_{0,1} e^{i\vec{k_1}\cdot\vec{r}}N_{\parallel1} \right)  \nonumber \\ \times \delta(\omega-v_\parallel k_{\parallel1}) \left(E_{\parallel 2} J_{0,2} e^{i\vec{k_2}\cdot\vec{r}}N_{\parallel2} \right)  \Bigg\}.
\end{eqnarray}
This diffusion coefficient parameterizes electron Landau damping by the LH wave. Using \eqref{eq:D2B} to rewrite $D_{ql,\parallel}$ in terms of the CQL3D quasilinear diffusion coefficients and applying the TORLH discretization \eqref{eq:torlhdisc} yields:
\begin{eqnarray}\label{eq:DQL_TORLH1}
    \textrm{B}(\psi,\theta,u_\parallel,u_\perp) &=& \frac{\pi e^2 [E]^2}{2 m_e^2 c^2}\textrm{Re}\Bigg\{\sum_{m_1,m_2} \frac{u_\parallel^4}{\gamma^2} \left(E_\parallel^{(m_2)} J_{0,2}e^{im_2\theta}N_\parallel^{(m_2)} \right)^* \nonumber \\
    && \times \delta(\omega - v_\parallel k_{\parallel}^{(m_1)}) \left(E_\parallel^{(m_1)} J_{0,1}e^{im_1\theta}N_\parallel^{(m_1)} \right) \Bigg\},
\end{eqnarray}
where $[E]$ is the electric field normalization in TORLH of 1 V/m. To efficiently discretize this problem in velocity-space we take advantage of the properties of the $\delta$ function. The $\delta$ function prescribes that for a given $m$ number the quasilinear diffusion coefficient has a uniquely defined $v_\parallel$ location in velocity-space. Therefore, at a given value of $m$ and $u_\perp$ the $u_\parallel$ velocity-space location is known. This allows us to rewrite (\ref{eq:DQL_TORLH1}) in terms of a discretized delta function:
\begin{eqnarray}\label{eq:DQL_TORLH2}
    \textrm{B}(\psi,\theta,m_1\propto u_\parallel,u_\perp) &=& \frac{\pi e^2 [E]^2 c^2}{2 \omega m_e^2}\textrm{Re}\Bigg\{\sum_{m_1,m_2} \frac{\gamma^2}{\left(N_\parallel^{(m1)}\right)^5} \Big(E_\parallel^{(m_2)} J_{0,2}e^{im_2\theta}N_\parallel^{(m_2)} \Big)^*  \nonumber \\
    && \times \frac{1}{\Delta u_{mesh}}\left( \frac{\partial v_\parallel}{\partial u_\parallel}\right)^{-1} \Big(E_\parallel^{(m_1)} J_{0,1}e^{im_1\theta}N_\parallel^{(m_1)} \Big) \Bigg\}.
\end{eqnarray}
Here we have used a rectangular discrete $\delta$ function where $\Delta u_{mesh}$ is the velocity-space mesh spacing. It is possible to use more complicated form factors for the discrete $\delta$, but these have been found to have a minimal impact on $D_{ql}$ formulation in the past \citep{Lee2017}. For numerical simplicity, a uniform velocity-space mesh in $u_\perp$ and $u_\parallel$ is used and the $u_\parallel = \gamma v_{ph\parallel}$ corresponding to a given $m$ number is interpolated to the nearest neighbor point on the $u_\parallel$ mesh. 
\begin{figure}
  \centering
  \includegraphics[width=0.85\linewidth]{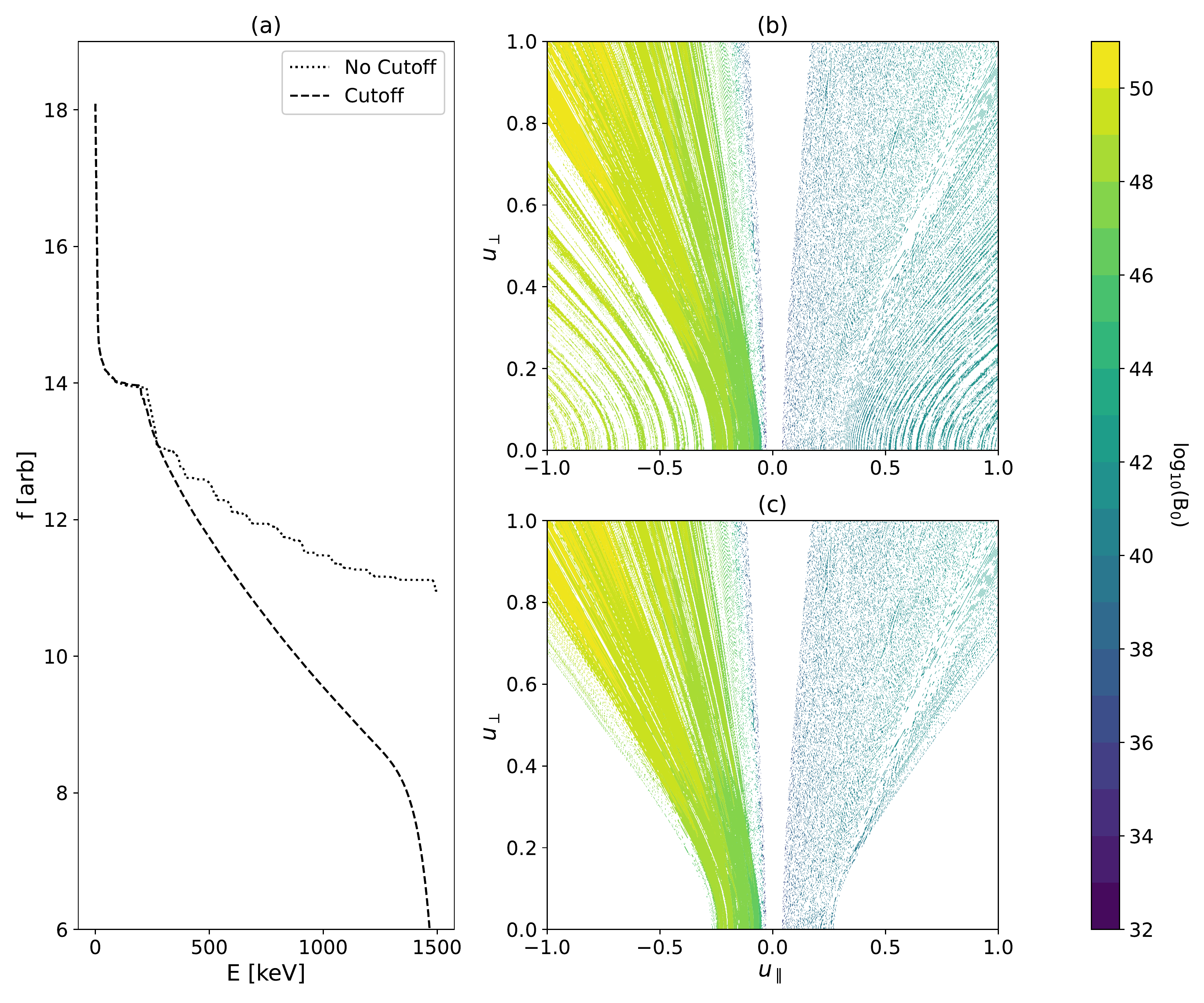}
  \caption{Plots demonstrating the evanescent cutoff noise reduction method. (a) The CQL3D electron distribution function produced by a $D_{ql}$ versus electron energy with (dashed) and without (dotted) the evanescence cutoff. Contours of $\textrm{B}_0$ versus parallel and perpendicular momentum per rest mass (b) without the cutoff applied and (c) with the cutoff applied.}
\label{fig:B0_cutoff}
\end{figure}

The $D_{ql}$ formulation here is effectively unchanged from that shown in \citet{Wright2014}, but a key numerical improvement to the $D_{ql}$ formulation has been made. Previously, a sum over all spectral modes, including evanescent wave-modes, was performed in TORLH. However, evanescent wave modes are disallowed in the resonant KE $D_{ql}$ formulation because waves with $\textrm{Im}[k_\parallel] \neq 0$ are incompatible with the contour integration used to obtain the resonant limit \citep{KennelEngelmann1966}. Inclusion of evanescent wave-modes in the $D_{ql}$ double mode sum introduces components of $D_{ql}$ at excessively high parallel momenta. This causes interaction with high-energy electrons leading to the formation of sharp structures in the FP solution. These sharp structures in phase-space caused numerical instabilities in both TORLH and CQL3D that could in some cases cause the simulations to fail. To avoid this, we check if the wave is evanescent using the following equation derived from the electromagnetic dispersion relation \citep{Bonoli1985}:
\begin{equation}\label{eq:cutoffcriterion}
    \big[(N_\parallel^2 - S)(\textrm{Re}\{\epsilon_\parallel (f)\} + S) + D^2\big]^2 - 4 \textrm{Re}\{\epsilon_\parallel (f)\}S(N_\parallel^2 - R)(N_\parallel^2 - L) \ge 0.
\end{equation}
For a given mode, if \eqref{eq:cutoffcriterion} is less than zero it is excluded from the double mode sum in \eqref{eq:DQL_TORLH2} (doing this calculation efficiently required a rewrite of the quasilinear diffusion coefficient formulation routine in TORLH to modify its parallelization). Ideally, we would solve the dispersion relation with fixed $N_\perp$ and $\omega$ for $N_\parallel$ to determine if the mode had $\textrm{Im}[k_\parallel]\neq 0$, but this is not possible in the TORLH basis. We must settle for calculating when $N_\perp$ becomes imaginary with fixed $N_\parallel$ and $\omega$. The results of this procedure, highlighted in Figure~\ref{fig:B0_cutoff}, show that it very successfully removes high energy noise from the $D_{ql}$ and CQL3D electron distribution function. 

To self-consistently obtain the desired power, the normalized TORLH electric fields in the $D_{ql}$ formulation must be scaled to the launched power in the experiment $P_{target}$. To do this, the launched power with the normalized field values in TORLH is obtained from the integrated Poynting flux at the antenna:
\begin{eqnarray}
    P_{RF,Poynt}(\psi_{ant}) &=& \frac{\pi l_0^2 [E]^2}{c \mu_0} \int_0^{2\pi} d\theta \frac{R_0}{l_0} n_\tau  \, \sum_{m,m^\prime} \textrm{Im}\Bigg\{e^{i(m-m^\prime)\theta} \nonumber \\ && \times \Bigg(E^{(m)*}_\eta \mathrm{R^{curl}_\zeta} \vec{E}^{(m^\prime)} - E^{(m)*}_\zeta \mathrm{R_\eta^{curl}} \vec{E}^{(m^\prime)} \Bigg)  \Bigg\},
\end{eqnarray}
where $\mathrm{R}^\textrm{curl}$ is the numerical curl operator in TORLH \citep{TORICManual}, $R_0$ is the major radius, and $l_0 = c/\omega$ is the TORLH length normalization. In the limit of only Landau damping, as in the simulations here, one can alternatively use the power deposited by Landau damping to scale the $D_{ql}$:
\begin{align}\label{eq:powLD}
    P_{RF,LD} &=  \frac{1}{2} \int dV \textrm{Re}\left\{\vec{E}^* \cdot \vec{J}_p\right\} \notag \\ &= - \pi \omega \epsilon_0 [E]^2 l_0^3 \int_0^1d\psi \sum_{m^\prime} \sum_m \textrm{Im}\left\{E_\parallel^{(m^\prime)*}(\psi)\tilde{\chi}_{zz}(m^\prime, \psi, \theta) E_\parallel^{(m)}(\psi) \right\},
\end{align}
where $\tilde{\chi}_{zz}$ is the Fourier transform of the imaginary component of the parallel dielectric:
\begin{equation}
\tilde{\chi}_{zz} = \int d\theta \mathcal{J}_n e^{i(m^\prime - m)\theta} \textrm{Im}[\chi_{zz}].
\end{equation}
Using the power calculated in TORLH we may rescale the quasilinear diffusion coefficient:
\begin{equation}
\textrm{B} = \textrm{B} \frac{P_{target}}{P_{RF}}.
\end{equation}
For linear field response, $P_{RF} \propto \textrm{B} \propto |E|^2$ making this rescaling physically self-consistent. No further rescaling is necessary, and precise agreement should be obtained between $P_{target}$ and the power calculated in CQL3D using the quasilinear diffusion coefficient.

A key feature of the quasilinear diffusion coefficients calculated by TORLH is their non-positive-definiteness. The holes in the diffusion coefficients shown in Figure~\ref{fig:B0_cutoff} are the result of small non-positive definite components of the $D_{ql}$ that are set to zero after output from TORLH. Non-positive-definite quasilinear diffusion is numerically unstable and physically incorrect, but is not unique to TORLH and is a weakness of spectral $D_{ql}$ reconstruction in many full-wave RF solvers \citep{Jaeger2006,Lee2017,LeePPCF2017}. The root cause of the non-positive-definiteness is improper consideration of parallel magnetic field inhomogeneity during the bounce average \eqref{eq:bounceavg} (the improper consideration of parallel magnetic field inhomgeneity also is the root cause of the issues related to evanescent modes discussed earlier). There are a number of ways to handle non-positive-definiteness: the negative components of $\textrm{B}_0$ may be set to zero and the diffusion coefficient normalized to conserve power (this is done automatically in CQL3D when given a non-positive-definite $D_{ql}$), an advanced positive definite spectral formulation may be used \citep{LeePPCF2017}, or a a particle tracking velocity kick formulation of $\textrm{B}_0$ can be performed \citep{pcommBH,Shiraiwa2011}. Advanced spectral formulations or particle tracking, however, are $\mathcal{O}(10^6)$ more computationally expensive than the traditional spectral $D_{ql}$ reconstruction technique described here or the W-dot method from \citet{Jaeger2006}. Positive definite $D_{ql}$ formulations also have unfavorable performance scaling with problem size quickly becoming more expensive than the electric field solve. While more sophisticated $D_{ql}$ formulation techniques may be applicable in less computationally expensive simulations of ion cyclotron heating \citep{LeePPCF2017}, their applicability to spectral LHCD calculations is limited.

\begin{figure}
  \centering
  \includegraphics[width=\linewidth]{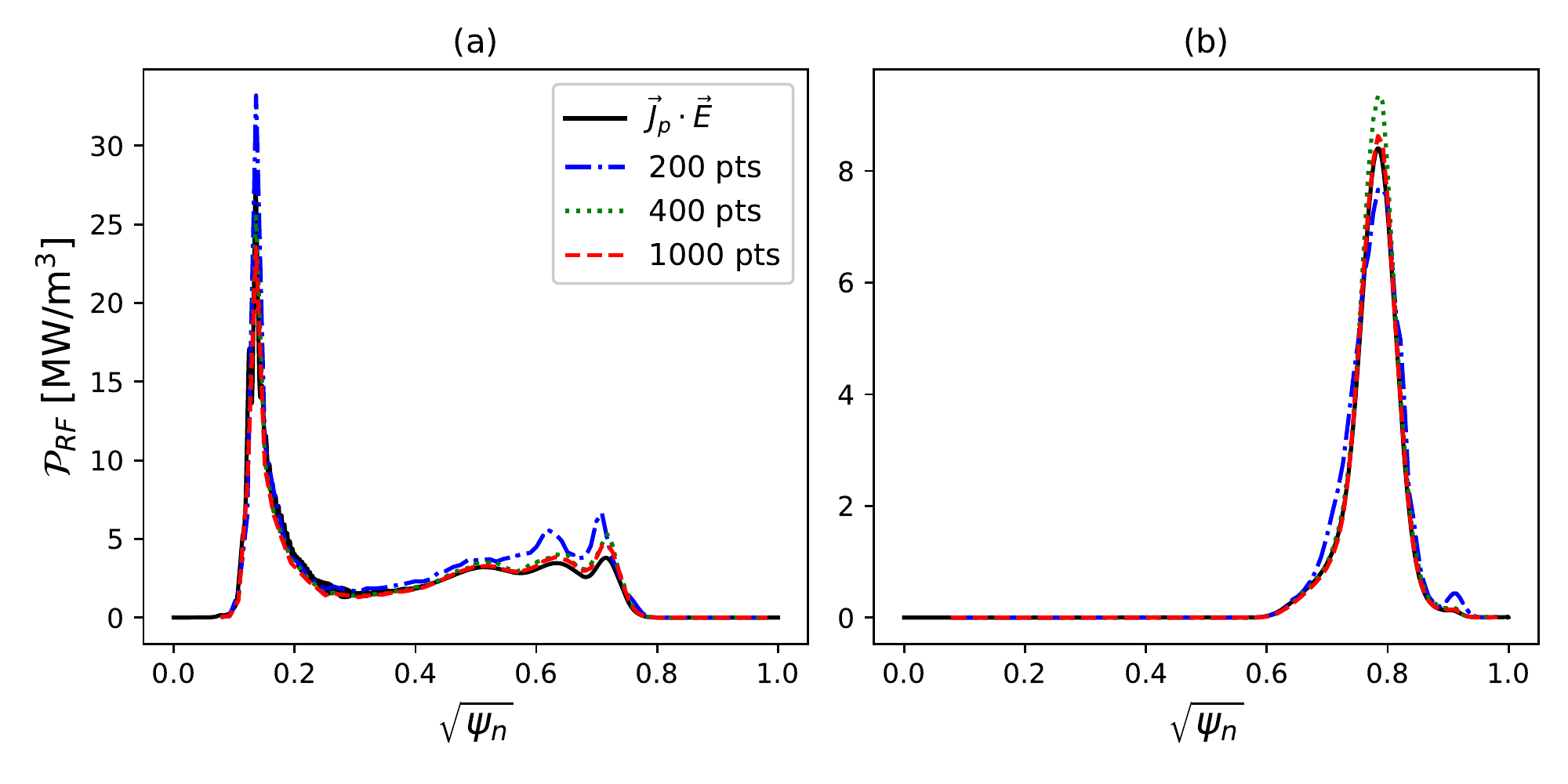}
  \caption{Plots of RF power density $\mathcal{P}_{RF}$ versus square-root poloidal flux $\sqrt{\psi_N}$ demonstrating velocity-space convergence of the $D_{ql}$. With increasing numbers of points in parallel velocity-space, $\mathcal{N}_{u\parallel}$, convergence improves. Note, for both plots here $\mathcal{N}_{u\parallel}=2\mathcal{N}_{u\perp}$. In (a) waves are weakly damped in an $N_\parallel=-1.6$ simulation, and in (b) waves are strongly single-pass damped in an $N_\parallel=-5.0$ simulation.}
\label{fig:dql_conv}
\end{figure}

Previously, 200-400 parallel velocity grid points were used in iterated simulations of TORLH and CQL3D \citep{Wright2014,Yang2014}. However, after the rewrite of the $D_{ql}$ formulation performed in this work, the velocity-space convergence requirements of the $D_{ql}$ were reanalyzed. To do this, we performed both standalone Maxwellian validation and iterated simulations with CQL3D. For brevity we will focus on the standalone validation here. For a given distribution function the RF power density $\mathcal{P}_{RF}$, calculated from power dissipation by integrated Landau damping on a flux surface $\vec{J}_p \cdot \vec{E}$ (integrated in $\psi$ in \eqref{eq:powLD}), may be related the second moment of the quasilinear diffusion term:
\begin{equation}\label{eq:jdotevqlflux}
    \mathcal{P}_{RF}(\psi) = \int\int d^3\vec{u} (\gamma-1)m_e c^2 Q(f).
\end{equation}
Using the CQL3D quasilinear diffusion term \eqref{eq:qlflux}, integrating by parts, and substituting a relativistic Maxwellian, $f=n_e\left(\frac{m_e}{2\pi k_B T_e} \right)^{3/2} e^{\frac{-2c^2(\gamma-1)}{v^2_{the}}}$ yields:
\begin{equation}\label{eq:powdql}
    \mathcal{P}_{RF} = \frac{4 n_e m_e}{\sqrt{\pi}v_{the}^5 \oint B_0 dl/B} \int_{-\infty}^\infty du_{\parallel 0} \int_0^\infty du_{\perp 0} u_{\perp 0} \frac{\textrm{B}_0}{\gamma^2} e^{\frac{-2c^2(\gamma-1)}{v^2_{the}}}.
\end{equation}
A comparison of the RF power density profiles obtained from (\ref{eq:jdotevqlflux}) allows us to evaluate the convergence of the quasilinear diffusion coefficient formulation in velocity-space. As the velocity-space grid resolution is increased the agreement between the two $\mathcal{P}_{RF}$ calculations should improve. Convergence of the $D_{ql}$ typically occurs when $\sim500$ to $1000$ parallel velocity grid points are used. An example of such a convergence scan can be found in Figure~\ref{fig:dql_conv}. In more strongly damped, high $N_\parallel$, problems the resolution requirement for $D_{ql}$ tends to be larger as power is concentrated in a smaller region of velocity-space. When the $D_{ql}$ is under-resolved, power deposition is over-predicted as poorly defined $D_{ql}$ edges cause excess power to be diffused into the distribution function. In these convergence scans we also studied the impact of the evanescent mode cutoff on power deposition profiles. We found adding or removing the evanescent mode cutoff to had no meaningful effect on the power deposition profiles obtained by integrating the quasilinear term. This is perhaps unsurprising as waves which are cutoff according to \eqref{eq:cutoffcriterion} have very short evanescence lengths and should not have the opportunity to deposit significant amounts of power in the plasma. Usually a higher resolution than that which is required for convergence of the Maxwellian problem is used for integrated non-Maxwellian simulations as edge effects will be exacerbated when a Landau plateau begins to form. Throughout this work, we use $1000$ points in the parallel velocity-space grid as convergence scans showed it consistently provided good convergence for both the non-Maxwellian and Maxwellian problem. 

Finally, unlike older versions of TORLH and TORIC \citep{Wright2010,Wright2014,Lee2017}, the $\textrm{B}_0$ calculated in TORLH here was passed to CQL3D directly. Exact flux surface matching between the two codes was enforced and a number of intermediate velocity-space interpolations which were present in previous work have been removed. This change markedly improved agreement between the $\mathcal{P}_{RF}$ profiles in CQL3D calculated using the quasilinear diffusion and TORLH calculated using $\vec{J}_p \cdot \vec{E}$ in integrated simulations.

\subsection{$\textrm{Im}[\chi_{zz}]$ Lookup Table Construction}\label{sec:chizzlookup}

In order to perform non-Maxwellian simulations of LHCD using TORLH it is necessary to calculate the non-Maxwellian dielectric response. In the case of LH waves this process is straightforward as only the imaginary parallel component of the dielectric $\textrm{Im}[\epsilon_\parallel] = \textrm{Im}[\chi_{zz,e}]$, which governs Landau damping, undergoes substantial changes (for plateau distributions resulting from Landau damping in the LH limit). However, calculating $\textrm{Im}[\chi_{zz,e}]$ accurately using realistic fully-relativistic distribution functions from CQL3D would be too slow to be performed during the dielectric construction in TORLH simulations. Instead a look up table, similar to those used for complicated dielectrics in AORSA and TORIC \citep{Berry2016,Bertelli2017}, is constructed which provides values of $\textrm{Im}[\chi_{zz,e}]$ on a fixed parallel refractive index $N_\parallel$, radial location $\psi$, and flux surface angle $\theta$ grid. $\textrm{Im}[\chi_{zz,e}]$ may be rapidly interpolated from the lookup table to avoid computational and memory bottlenecks during the dielectric construction step in TORLH.

The derivation of $\textrm{Im}[\chi_{zz,e}]$ in terms of $N_\parallel$, $\psi$, and $\theta$ for a CQL3D distribution function is performed in \citet{Wright2010} and reviewed in Appendix~\ref{sec:chizzform}. However, an overlooked aspect of implementing this $\textrm{Im}[\chi_{zz}]$ formulation was the method by which the derivatives of the CQL3D distribution were taken. Originally, derivatives of the CQL3D distribution were taken using an uncentered $1^{st}$ order upwind difference. However, verification tests of TORLH performed during this work revealed, when a $\textrm{Im}[\chi_{zz,e}]$ produced from a Maxwellian distribution with these derivatives was used in TORLH, poor agreement was obtained with the same simulations using analytic expressions for the Maxwellian $\textrm{Im}[\chi_{zz,e}]$. 

Initially, discrepancies in the Maxwellian regression tests were attributed to the relativistic correction to $\textrm{Im}[\chi_{zz}]$s derived numerically from CQL3D distribution functions (the canonical analytic Maxwellian $\textrm{Im}[\chi_{zz,e}]$ in TORLH using the $\textrm{Z}$ function is non-relativistic), but iterated simulations of TORLH and CQL3D also demonstrated poor agreement between TORLH and CQL3D damped powers. The root cause of the disagreement was found to be a small problem with the derivative method used. In rapidly varying functions, like a plasma distribution function at large $u_\parallel$, it is insufficient to use un-centered differences. Re-centering of the grid at the derivatives' locations is important to obtaining accurate results. This is evident in CQL3D where great care is taken to properly recenter differences \citep{pcommBH}. To demonstrate this we introduce analytic Maxwellian:
\begin{subeqnarray}
    \frac{\partial F(w)}{\partial w} &=& -2Cw e^{w^{-2}}\label{eq:dervMax} \\
    F(w) &=& C e^{w^{-2}}, \label{eq:normMax}
\end{subeqnarray}
and a 1-D Landau plateau distribution function obtained by solving the Fokker-Planck equation with quasilinear RF diffusion for the Trubinkov collision operator \citep{Trubnikov1965,Fisch1978,Karney1979}:
\begin{subeqnarray}
    \frac{\partial F(w)}{\partial w} &=&  - \frac{2F(w)}{w^2(2D(w)+1/w^3)}\label{eq:dervpiece} \\
        F(w) &=& C \exp \left[ -\int^w \frac{2w \ dw}{1+2w^3D(w)} \right],
\end{subeqnarray}
for:
\begin{equation}
    D(w) = 
    \begin{cases}   D_0 & w_1 \leq w \leq w_2 \\ 
    0 & \ \ elsewhere 
    \end{cases}
\end{equation}
For this choice of a $D(w)$, \eqref{eq:dervpiece} has closed form solution (not previously noted in the literature):
\begin{equation} \label{eq:fpiece}
  F(w) = 
    \begin{cases}
	C_1 \exp(-w^2) & w < w_1 \\
	C_2 \Bigg[ \frac{\left( 1 + D_0^{1/3}w \right)^2}{ 1 - (2 D_0)^{1/3}w + (2 D_0)^{2/3}w^2 } \Bigg]^{1/6(2D_0)^{2/3}} \\ \quad \quad \quad \times \exp \left[-\frac{2\sqrt{3}}{6(2D_0)^{2/3}} \arctan\left( \frac{-1+2D_0^{1/3}w}{\sqrt{3}} \right) \right] & w_1 \leq w \leq w_2 \\
	C_3 \exp(-w^2) & w_2 < w,
	\end{cases}
\end{equation}
where $C_1$ is an arbitrary normalization constant and constants:
\begin{subeqnarray}
    C_2 = C_1 \left[ \frac{1 - (2 D_0)^{1/3}w_1 + (2 D_0)^{2/3}w_1^2}{\left( 1 + D_0^{1/3}w_1 \right)^2}\right]^{1/6(2D_0)^{2/3}}\\
    \times \exp \left[-\frac{w_1^2}{2}+\frac{2\sqrt{3}}{6(2D_0)^{2/3}} \arctan\left( \frac{-1+2D_0^{1/3}w_1}{\sqrt{3}} \right) \right]\nonumber\\
    C_3 = C_2 \left[ \frac{\left( 1 + D_0^{1/3}w_2 \right)^2}{ 1 - (2 D_0)^{1/3}w_2 + (2 D_0)^{2/3}w_2^2 }\right]^{1/6(2D_0)^{2/3}}\\
    \times \exp\left[\frac{w_2^2}{2} - \frac{2\sqrt{3}}{6(2D_0)^{2/3}} \arctan\left( \frac{-1+2D_0^{1/3}w_2}{\sqrt{3}} \right) \right] \nonumber
\end{subeqnarray}
\begin{figure}
  \centering
  \includegraphics[width=1.0\linewidth]{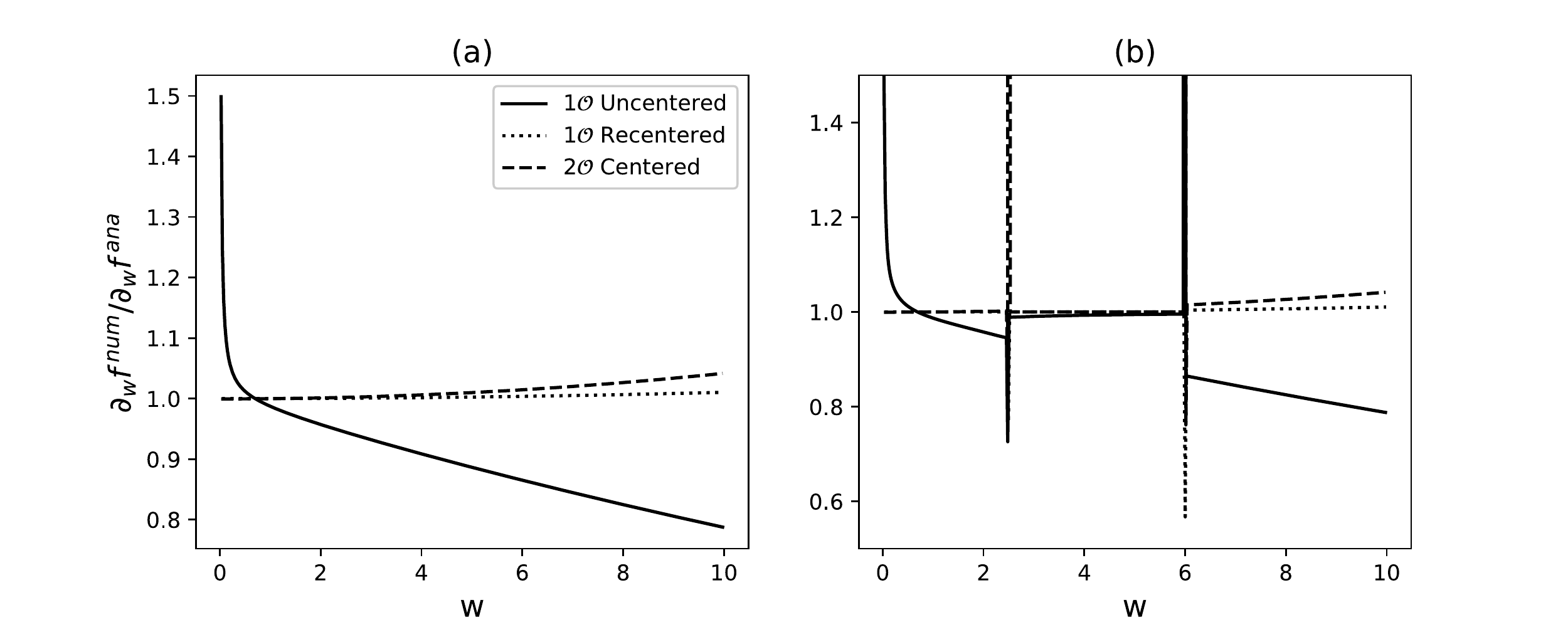}
  \caption{The ratio of the analytic and numerical derivatives taken using a $1^{st}$ order upwind forward difference (solid), a re-centered $1^{st}$ order upwind forward difference (dotted), and a $2^{nd}$ order central difference (dashed), for (\textit{a}) the Maxwellian distribution from \eqref{eq:normMax} and (\textit{b}) the non-Maxwellian distribution function from \eqref{eq:fpiece}.}
\label{fig:dervcompare}
\end{figure}
Both of these equations (\ref{eq:normMax}b) and \eqref{eq:fpiece} have well defined analytic derivatives (\ref{eq:dervMax}a) and \eqref{eq:dervpiece}, except at points on the edges of $D(w)$ where the \eqref{eq:fpiece} derivatives are undefined. If we ensure our numerical grid does not intersect with the edges of $D(w)$, we can use these equations to precisely evaluate the effectiveness of finite-differencing methods. We show a comparison of finite difference methods: $1^{st}$ order uncentered upwind, $1^{st}$ order upwind with grid re-centering, and $2^{nd}$ order central differences in Figure~\ref{fig:dervcompare}. Without re-centering, $1^{st}$ order differences perform extremely poorly explaining the difficulties experienced with the original $\textrm{Im}[\chi_{zz,e}]$ calculation. However, when re-centering is applied $1^{st}$ order upwind differences offer comparable performance to $2^{nd}$ order central differences, but unlike higher order difference, cannot produce the wrong derivative sign. Recentered $1^{st}$ order upwind differences are used in CQL3D, and iterated simulation tests produced the most robust agreement between TORLH and CQL3D when $1^{st}$ order upwind differences were also used in the $\textrm{Im}[\chi_{zz,e}]$ calculation (even when compared to $3^{rd}$ and $4^{th}$ order accurate differencing schemes). Presumably, this was the result of increased internal self-consistency. The correction to the derivatives in the $\textrm{Im}[\chi_{zz,e}]$ calculation immediately improved the agreement between the damped power calculations in TORLH and CQL3D and this correction was key to enabling the work in the following sections.

\section{Quasi-Steady State Simulation}\label{sec:steadystate}

The first tokamak discharge modeled here is Alcator C-Mod shot \#1060728011. In this discharge LHCD was produced with the Alcator C-Mod LH1 antenna at 4.6 GHz with 60$^{\circ}$ phasing corresponding to a launched $N_\parallel = -1.6$ \citep{ShiraiwaNF2011}. Approximately 800 kW of LHCD power was coupled to a plasma with $T_{e0} = 2.3$ keV, $\bar{n}_e = 5 \times 10^{19}$ m$^{-3}$ producing a near non-inductive plasma operating at $I_p = 540$ kA with a loop voltage of $-0.2$ V. This discharge has been modeled extensively in the past using both raytracing \citep{SchmidtThesis} and full-wave simulations \citep{Wright2009,Wright2010,Wright2014,MeneghiniThesis}. Furthermore, this shot is very similar to the shot \#1101104011 that has also been subject to much analysis in the literature \citep{MumgaardThesis,Biswas2020,Baek2021,Biswas2021}.

\begin{figure}
  \centering
  \includegraphics[width=0.75\linewidth]{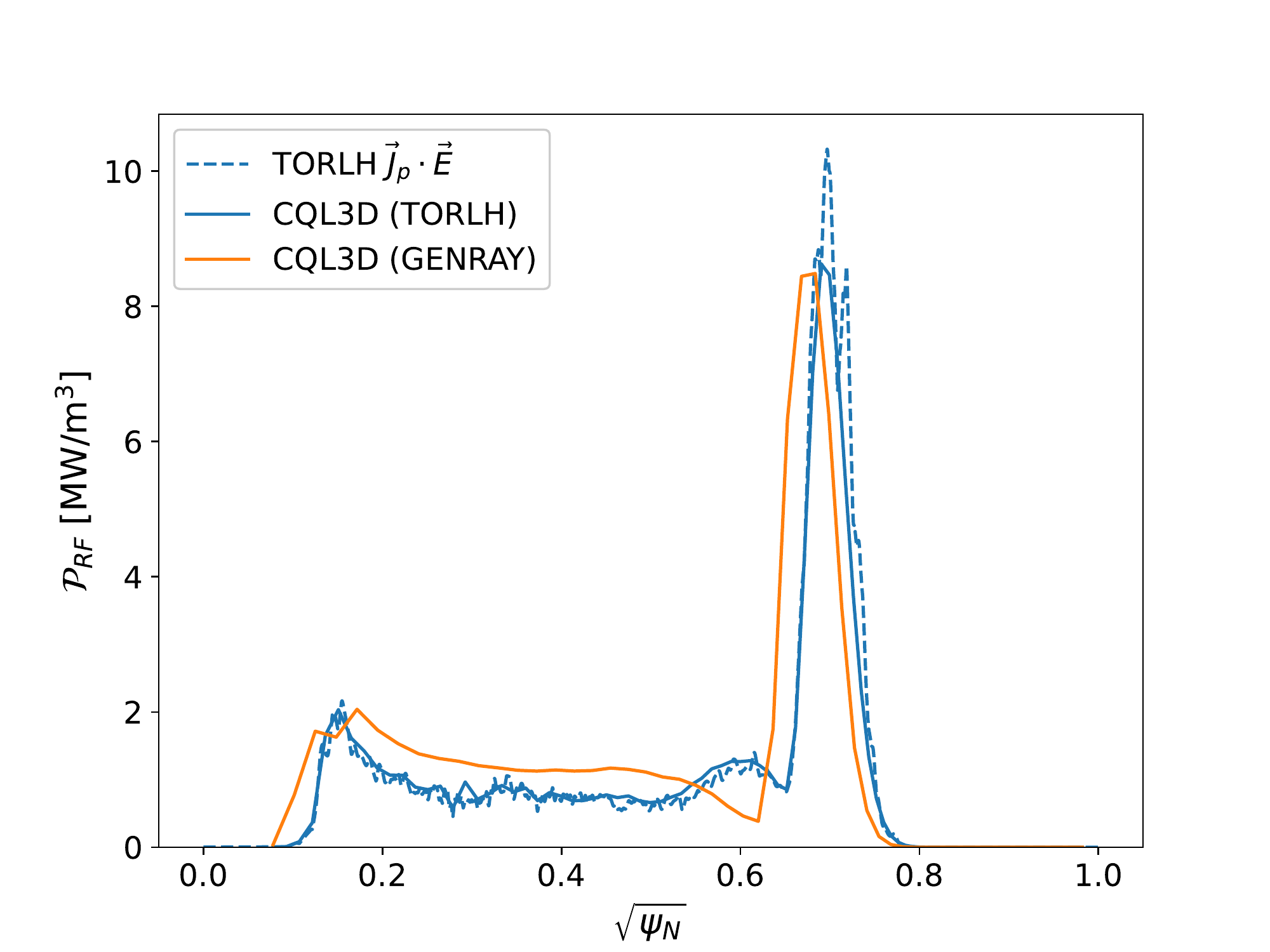}
  \caption{RF power density $\mathcal{P}_{RF}$ versus square-root poloidal flux $\sqrt{\psi_N}$ at the final iteration of a TORLH/CQL3D (blue) and a GENRAY/CQL3D (orange) simulation of Alcator C-Mod shot \# 1060728011. Lines corresponding to the RF power density from integration of the CQL3D quasilinear diffusion term are solid and lines corresponding to RF power density obtained from integration of the TORLH $\vec{J}_p\cdot\vec{E}$ are dashed.}
\label{fig:pwr10728011}
\end{figure}

Here we return to shot \#1060728011 using an overhauled TORLH and improved methodology to reevaluate previous results. As in \citet{Frank2022}, we have used the Integrated Plasma Simulator \citep{Elwasif2010} to create matched raytracing simulations with plasma state files from the full-wave simulation to evaluate the importance of full-wave effects. The GENRAY raytracing simulations here used the same source codes as those in \citet{Frank2022} in which the dispersion relation is based on the TORLH dielectric and includes hot plasma corrections to the real part of the dispersion relation. The GENRAY simulation of this shot used 400 rays, all with $N_\parallel = -1.6$, spread evenly over four grills placed at $\pm$ 30$^{\circ}$ and $\pm 10^{\circ}$ from the outboard midplane. The raytracing simulation results were coupled to a CQL3D simulation that used 60 flux surfaces equi-spaced from $\sqrt{\phi_n}$ = 0.05 to 0.95, where $\phi_n$ is the normalized toroidal flux. The distribution function in the CQL3D simulation was evolved for $\sim125$ ms over 35 timesteps. Timestep size was progressively increased from 1 $\mu$s to 10 ms during the simulation for numerical stability (while the time advance of the FP equation in CQL3D is implicit, the updates to the quasilinear diffusion coefficient when CQL3D is coupled with GENRAY are explicit and can become numerically unstable if long timesteps are used initially).
\begin{figure}
    \centering
    \includegraphics[width=0.425\linewidth]{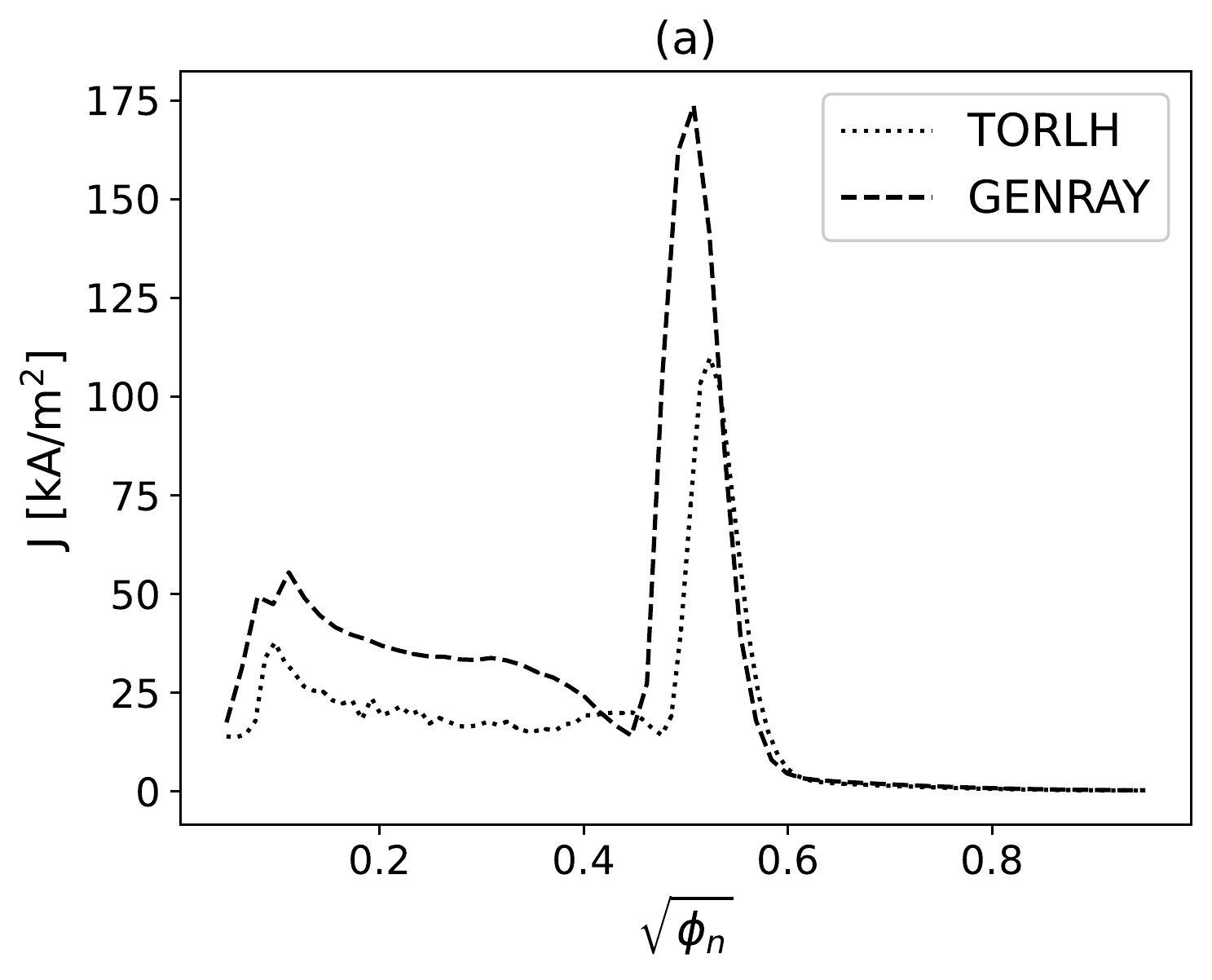}
    \includegraphics[width=0.525\linewidth]{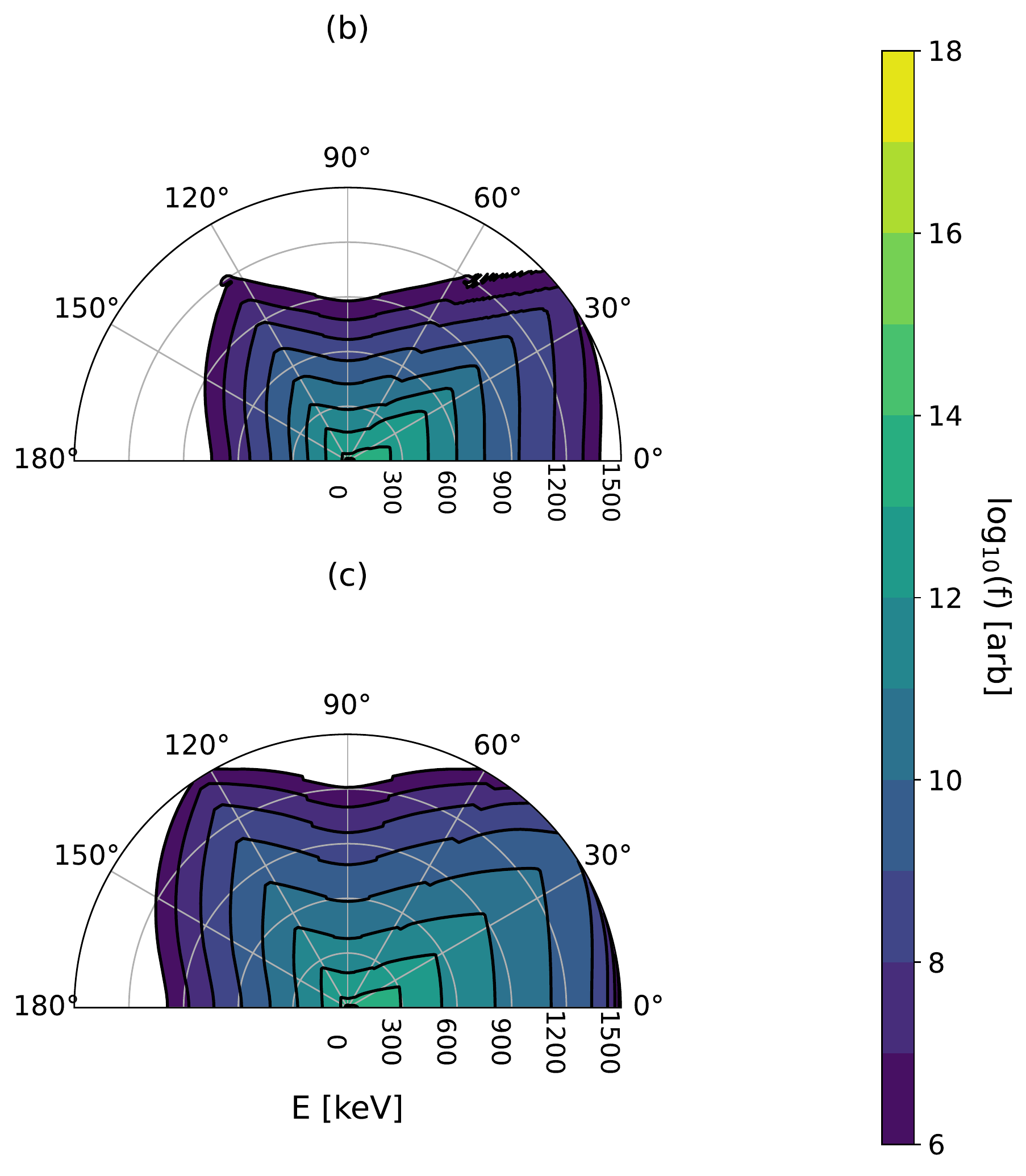}
    \caption{(a) RF driven current density $J$ versus normalized square-root toroidal flux $\sqrt{\phi_n}$ in a TORLH/CQL3D simulation (dashed) and a GENRAY/CQL3D simulation (dashed). Contours of the electron distribution function $f$ at $\sqrt{\phi_n}\sim0.5$ versus $E$ in keV, and velocity-space pitch angle $\vartheta$, for (b) a TORLH/CQL3D simulation, and (c) a GENRAY/CQL3D simulation. Both plots (b) and (c) use the same contour levels to more effectively show the differences in RF diffusion of power to electrons with high energies.}
    \label{fig:2ddist}
\end{figure}

The TORLH simulations of this discharge were performed using 2047 poloidal modes and 4800 radial finite elements and were run on 255 nodes and 8160 cores on Cori supercomputer at NERSC. The TORLH resolution settings here follow the requirements detailed in \citet{Frank2022}, however, higher finite element resolution was needed to combat spectral pollution which could occur when strong radial discontinuities in the dielectric function for a fixed poloidal mode $m$ appeared due to Landau plateau formation. The boundary condition was matched to the raytracing simulations using the improved boundary condition in \citet{Frank2022}, and an $N_\parallel = -1.6$ was launched. The CQL3D simulations coupled to full-wave simulations used 96 flux surfaces equi-spaced over $\sqrt{\phi_n} = 0.05$ to 0.95. The larger number of flux surfaces in the Fokker Planck calculation relative to the raytracing/Fokker-Planck simulations was another measure taken to improve numerical stability of the non-Maxwellian TORLH simulations. The TORLH/CQL3D simulations used a ramped timestepping scheme similar to the raytracing/CQL3D simulations to improve numerical stability as TORLH and CQL3D are iterated explicitly within the IPS as described in Section~\ref{sec:torlhcql3dcoupling}. A total of 40 iterations between TORLH and CQL3D were performed here for a total FP time of $\sim 90$ ms. The TORLH and CQL3D simulation was well converged, TORLH power matched CQL3D power with discrepancies of $\lesssim 10\%$ without the use of numerical rescaling factors (the TORLH power target was 800~kW and the total damped power in CQL3D was 719~kW). 
\begin{figure}
  \centering
  \includegraphics[width=0.45\linewidth]{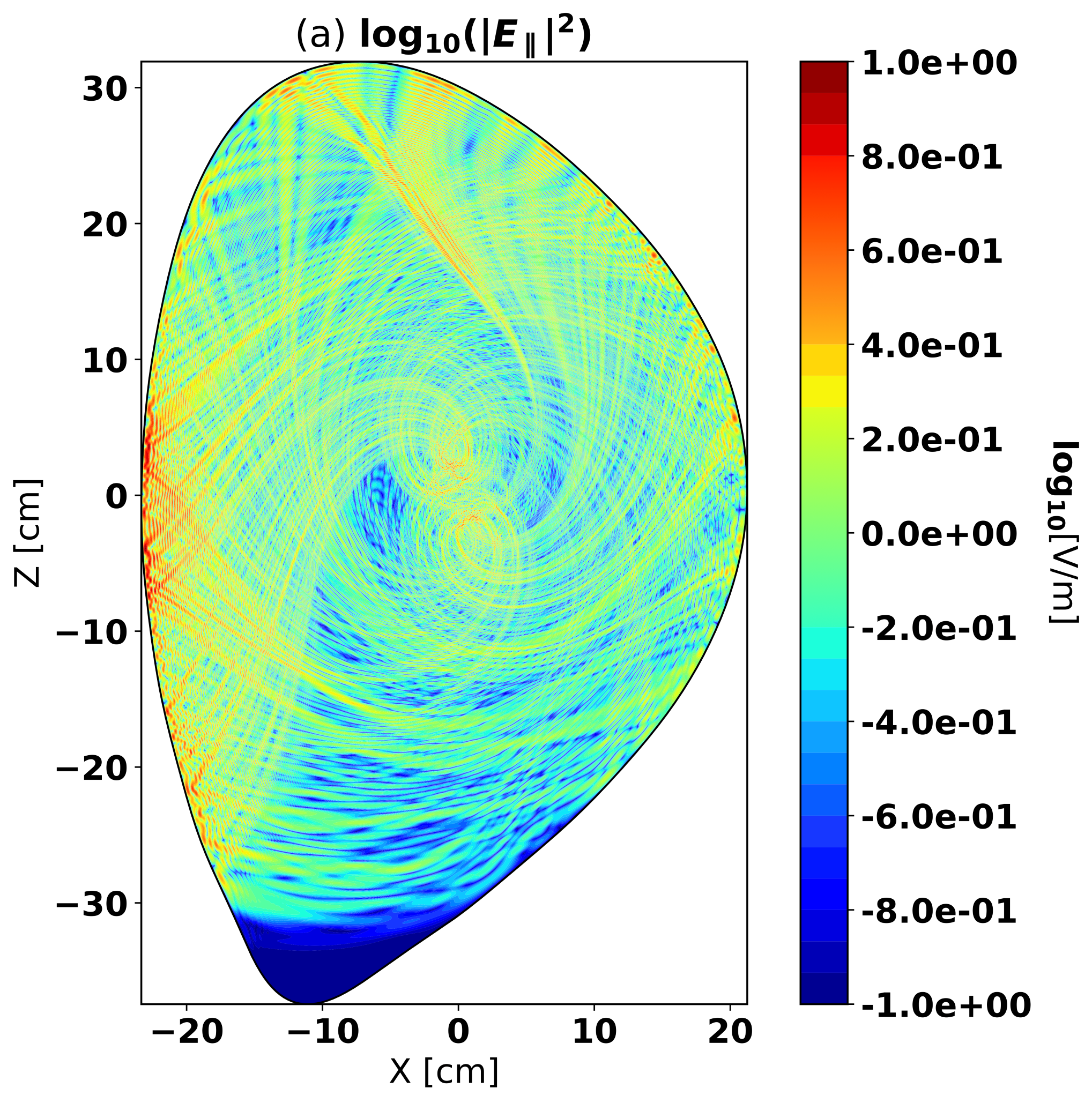}
  \includegraphics[width=0.45\linewidth]{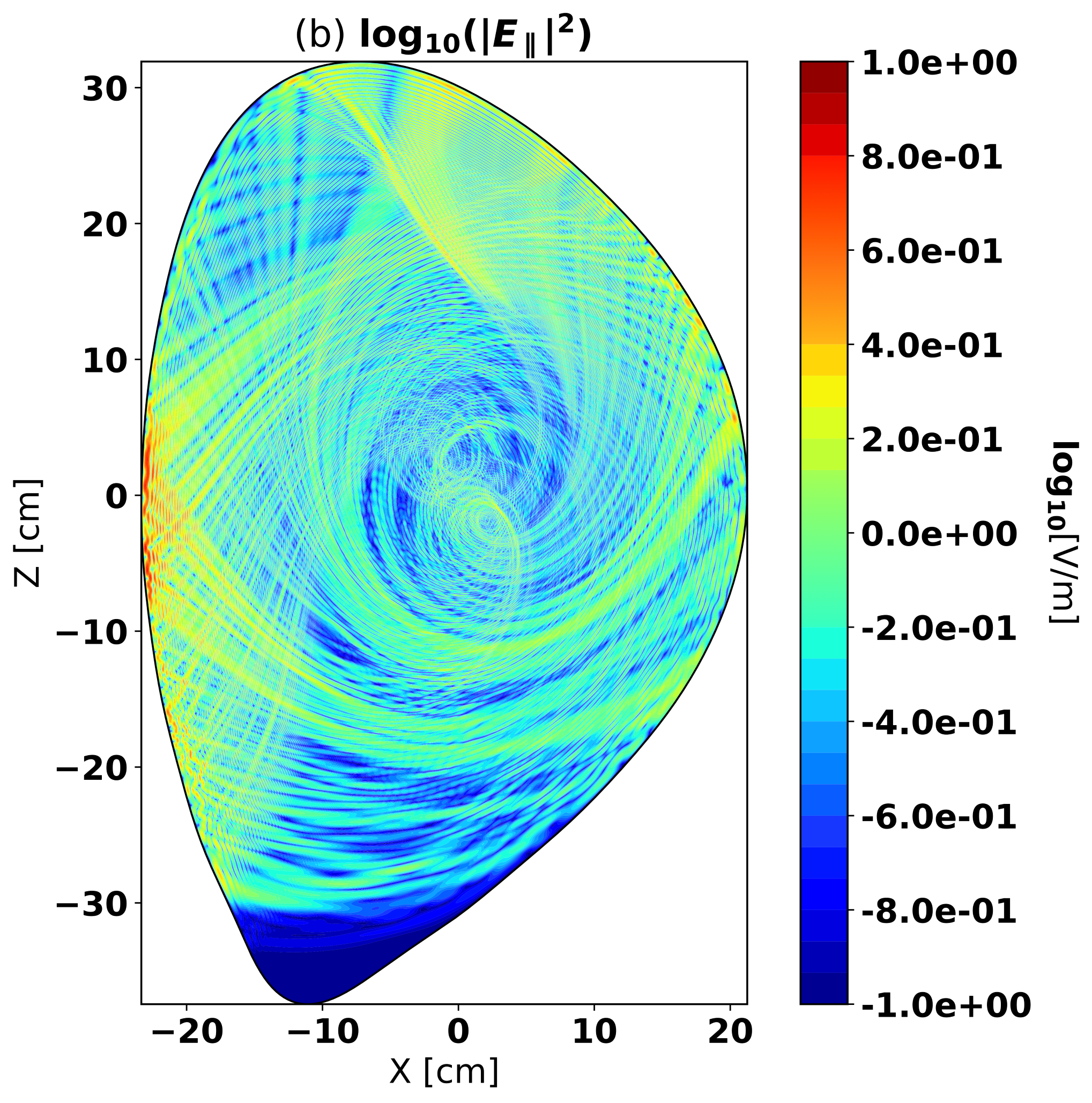}
  \caption{The logarithmic contours of the normalized TORLH electric field (launched field in TORLH is normalized to 1 V/m) in simulations of Alcator C-Mod shot \#1060728011 for Maxwellian damping (a), and non-Maxwellian damping (b).}
\label{fig:fields}
\end{figure}

Comparisons of the $\mathcal{P}_{RF}$ profiles from TORLH/CQL3D and GENRAY/CQL3D simulations, shown in Figure~\ref{fig:pwr10728011}, demonstrate that excellent power deposition agreement was obtained in the non-Maxwellian simulations, much like in the Maxwellian simulations of this discharge performed in \citet{Frank2022}. Power deposition profiles at the final iteration calculated by CQL3D from integration of the quasilinear diffusion term for both GENRAY and TORLH simulations closely agreed with one another and agreed with the integrated $\vec{J}_p\cdot\vec{E}$ calculated by the TORLH field solver. Despite excellent $\mathcal{P}_{RF}$ agreement, TORLH/CQL3D simulations had lower current drive efficiency than GENRAY/CQL3D simulations. Current profiles in both simulations closely matched the power deposition profiles, but the TORLH/CQL3D simulations drove 288 kA of current with 719 kW of input power compared to the GENRAY/CQL3D simulations which drove 455 kA of current with 800 kW of input power. The shortfall in the TORLH/CQL3D current drive efficiency, which is only $\sim70$\% of the GENRAY/CQL3D value, appears to be the result of holes in the quasilinear diffusion coefficient introduced by the non-positive definite nature of the coefficient, discussed in Section~\ref{sec:dql}. These holes limit the power which can be diffused to higher energy electrons. The discontinuities from the holes in the quasilinear diffusion coefficient cause a sharp drop in the distribution about the hole. This seems to largely be a 2D effect as the holes do not substantially affect the RF power deposition or current drive profiles, and the power deposition profile is know to be driven primarily by the 1D distribution function behavior \citep{Bonoli1986,MeneghiniThesis,Shiraiwa2011}. However, the quasilinear diffusion holes substantially reduce the net driven current which \textit{is} sensitive to changes to the 2-D distribution \citep{Fisch1980}. This is demonstrated in Figure~\ref{fig:2ddist} where it is shown that, despite having very similar current drive profiles, GENRAY/CQL3D simulations drive more current than TORLH/CQL3D simulations, and power is more effectively diffused to higher energy electrons in GENRAY/CQL3D simulations. We found processing the quasilinear diffusion coefficient produced by TORLH with a biharmonic in-painting algorithm \citep{Damelin2018,Chui2010} to remove the holes from non-positive definiteness increased the current drive efficiency substantially. However, in-painting also spoiled the power calculation's self-consistency and led to an unstable iteration between TORLH and CQL3D that could not be used.

Unlike previous non-Maxwellian simulations of this discharge in \citet{Wright2014} there was not as drastic a drop in the non-Maxwellian core electric fields at the first and the last iteration of the TORLH/CQL3D simulation, as shown in Figure~\ref{fig:fields}. The drop observed by \citet{Wright2014} was due to the physically inconsistent nature of their calculations and an error in the $\chi_{zz}$ table interpolation that lead to wave damping being over-predicted. Despite having increased field amplitudes in our simulation, constructive interference does not seem to cause substantial differences in the power deposition profiles versus GENRAY. This has a simple explanation; the damped power's dependence on the slope of the distribution, $\propto \exp(-\zeta^2)$ for a Maxwellian distribution or $\propto \partial f / \partial v_\parallel$ for a non-Maxwellian distribution, is much stronger driver of power deposition location in $\psi$ than the damped power's dependence on field intensity $\propto |E_\parallel|^2$. Thus, the power deposition profile should only be weakly affected by interference in most cases. Our result here demonstrates that the GENRAY simulations of weakly damped LHCD, which have been found to differ from experiment \cite{SchmidtThesis,Wright2014,MumgaardThesis}, accurately reproduce the power deposition profiles calculated with full-wave simulations. This indicates that the discrepancy between raytracing/FP results and experimentally measured current profiles is \textit{not} due to full-wave effects such as diffraction or interference. Much like raytracing/FP, full-wave/FP simulations cannot accurately replicate the experimentally measured ohmic-like LHCD profiles which are peaked on axis.

\section{LH Modulation Scan Simulations}

\begin{figure}
  \centering
  \includegraphics[width=1\linewidth]{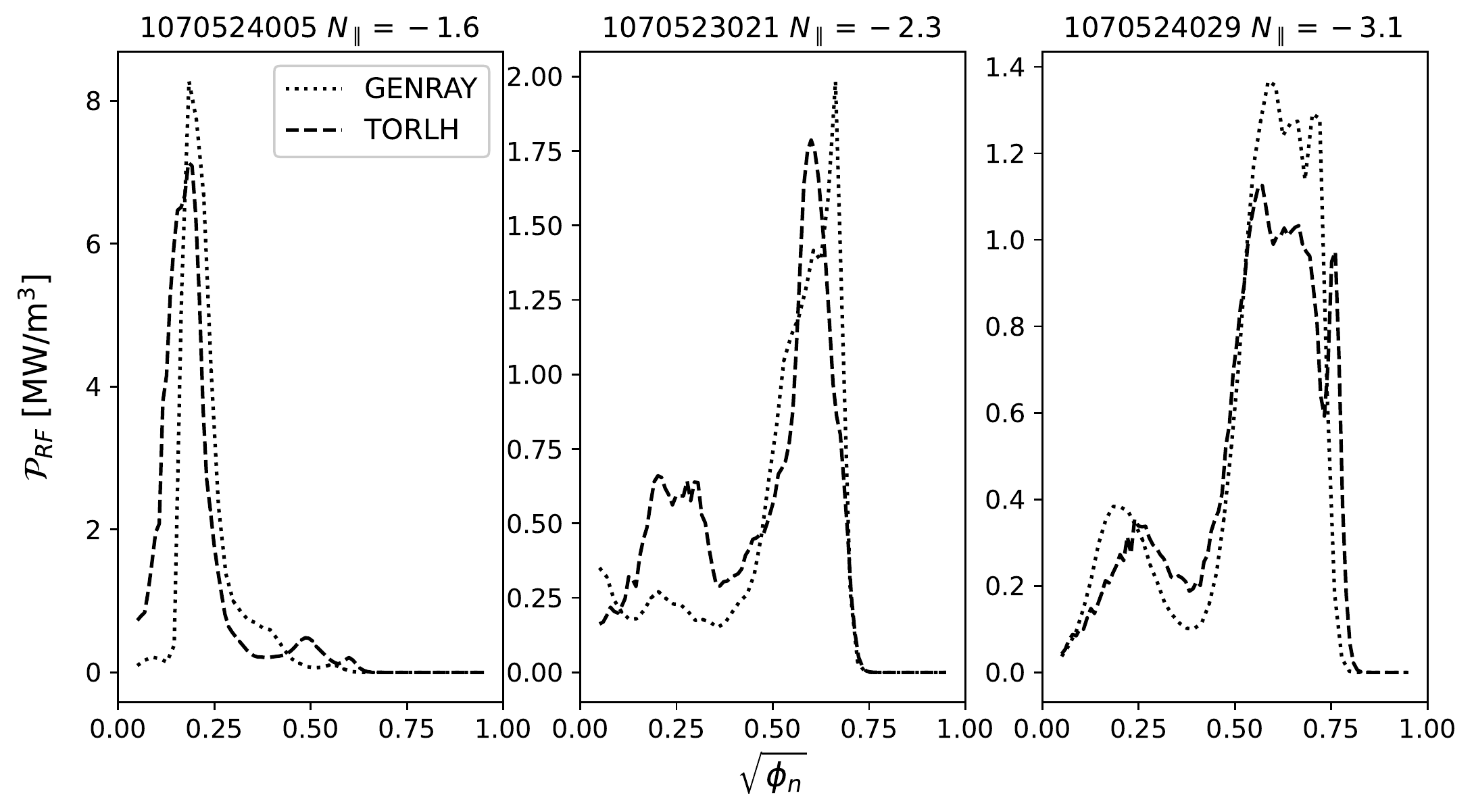}
  \caption{RF power density $\mathcal{P}_{RF}$ versus square-root toroidal flux $\sqrt{\phi}$ calculated by CQL3D coupled with GENRAY (dotted) or TORLH (dashed) for 3 different Alcator C-Mod discharges using different LHCD launch phasings.}
\label{fig:power2}
\end{figure}

The next set of Alcator C-Mod discharges modeled here were the LHCD modulation experiments performed by \citet{Schmidt2011}. In these experiments 400 kW of LHCD was modulated with a 50\% duty cycle and a pulse time of 12.5 ms, in discharges that were primarily ohmic ($V_{loop} = 1.0$ V) at high density $\bar{n} = 9.0 \times 10^{19}$ m$^{-3}$, with $T_{e0} \sim 2.2$ keV. Three different launcher phasings $60^\circ$, $90^\circ$, and $120^\circ$, corresponding to $N_\parallel = $ $-1.6$, $-2.3$, and $-3.1$ respectively were used in these experiments. GENRAY/CQL3D modeling was performed in the original analysis in \citet{Schmidt2011,SchmidtThesis} and compared to inverted experimental hard X-ray measurements with mixed results. Qualitative agreement with experiment was obtained when $N_\parallel = -2.3$ and $-3.1$, but agreement with experiment was very poor when $N_\parallel = -1.6$.

We repeated the analysis of these discharges with GENRAY/CQL3D and matched TORLH/CQL3D simulations using the IPS. The GENRAY simulations used 200 rays spread over 4 launch points to mimic the Alcator C-Mod LH1 launcher, the same custom source code described in Section~\ref{sec:steadystate}, and used a single fixed $N_\parallel$ value (except when otherwise noted). CQL3D in the GENRAY/CQL3D simulations used 48 equi-spaced flux surfaces and took 50, 0.25 ms timesteps for a total FP time of 12.5 ms (equivalent to the LHCD pulse time in the modulation experiments). The TORLH simulations used 2047 poloidal modes and 6000 radial finite elements and were run using 8160 cores and 255 nodes on Cori at NERSC in iterated simulations that lasted $\sim 36$ hrs. The larger finite element number than the simulations in Section~\ref{sec:steadystate} eliminated spectral pollution which occurred in the edge of the $N_\parallel = -2.3$ and $-3.1$ simulations. The CQL3D settings in the TORLH/CQL3D integrated simulations were the same as those in Section~\ref{sec:steadystate}, but CQL3D was iterated with TORLH only 25 times in order to achieve a FP-time of $12.55$ ms. All the simulations here used 400 kW of launched power like the experiments. The integrated $\vec{J}_p\cdot\vec{E}$ and power profiles obtained in CQL3D from integration of the quasilinear diffusion term closely agreed in all cases. Furthermore, the integrated power in all CQL3D simulations was within 2\% of the 400 kW target meaning the degree of self-consistency achieved in these simulations was higher than any previous full-wave Fokker-Planck modeling study \citep{Jaeger2006,Shiraiwa2011,Wright2014,Lee2017,LeePPCF2017,Bertelli2017}.

\begin{figure}
  \centering
  \includegraphics[width=0.65\linewidth]{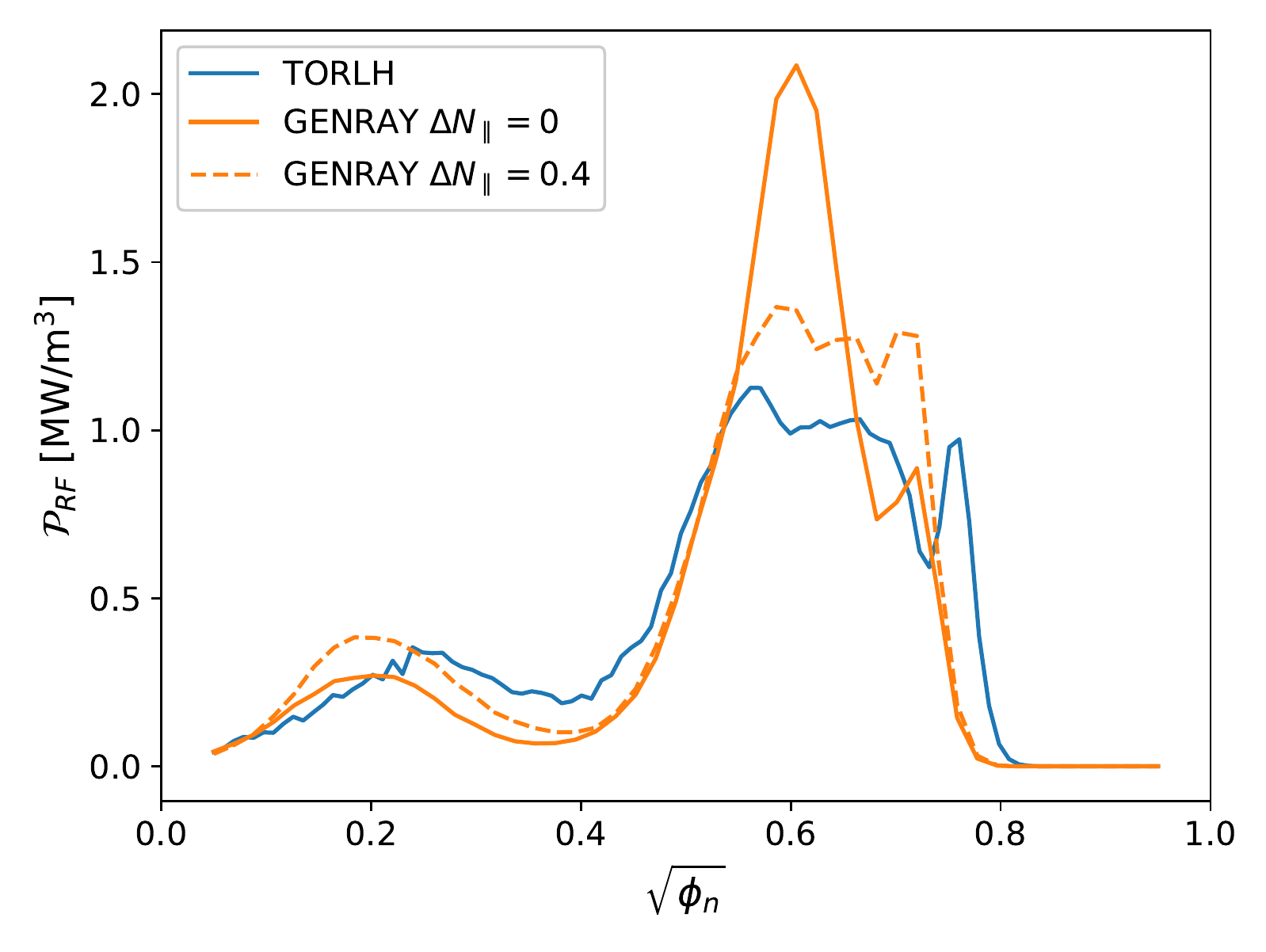}
  \caption{RF power density $\mathcal{P}_{RF}$ versus square-root toroidal flux $\sqrt{\phi}$ calculated by CQL3D coupled with GENRAY (blue) or TORLH (orange) for two different values of GENRAY spectral broadness $\Delta N_\parallel = 0$ (solid) and $\Delta N_\parallel = 0.4$ (dashed).}
\label{fig:deltanparcompare}
\end{figure}

The RF power deposition profiles obtained from these simulations, shown in Figure~\ref{fig:power2}, found once again that GENRAY/CQL3D accurately reproduced the power deposition profiles calculated using TORLH/CQL3D. However, in the $N_\parallel = -3.1$ case GENRAY/CQL3D initially predicted a narrower power deposition peak. To obtain a close RF power deposition profile match with TORLH/CQL3D we had to broaden the launched GENRAY spectrum. This difference almost certainly was due to the influence of diffraction in the TORLH simulation, as diffraction is expected to broaden the spectrum symmetrically \citep{Pereverzev1992}.  A scan of $\Delta N_\parallel$ values from 0.1 to 0.6 was performed and a GENRAY launcher spectrum with $\Delta N_\parallel = 0.4$ was found to be the best match to the TORLH results. A comparison of the unbroadened and broadened damping is found in Figure~\ref{fig:deltanparcompare}. The spectral broadening needed to account for diffraction is substantially less than the spectral width of the LH1 launcher in C-Mod where $\Delta N_\parallel = 1.0$ \citep{ShiraiwaNF2011}, indicating that the effects of diffractional broadening are relatively small. Furthermore, in the more weakly damped $N_\parallel =$ -2.3 and -3.1 cases no spectral broadening in GENRAY was needed to obtain good agreement with TORLH. This indicates that the geometric upshift effect \citep{Bonoli1981,Bonoli1982}, included in both simulations, is the dominant gap closure mechanism. Diffraction is only a small correction and relatively unimportant that can impart only modest broadening on the wave spectrum. Importantly, our results agreed well with the GENRAY/CQL3D results in \citet{Schmidt2011}, and neither the TORLH/CQL3D or GENRAY/CQL3D simulations agreed with experimental measurements in the weak damping, $N_\parallel = -1.6$, case where ohmic-like profiles were measured \citep{Schmidt2011}.

Finally, current drive calculations were performed in each case. As in Section~\ref{sec:steadystate}, we observed GENRAY/CQL3D predicted total driven currents fairly close to experiments, but the TORLH/CQL3D simulations systematically under-predicted the current drive as a result of the holes present in the quasilinear diffusion coefficient. Current drive profiles in both the GENRAY/CQL3D and TORLH/CQL3D simulations closely followed the power deposition profiles, however, the TORLH/CQL3D current drive efficiency was again roughly $\sim 70 \%$ the value obtained in by the GENRAY/CQL3D simulations. 

\section{Conclusion}

We performed the first fully self-consistent TORLH/CQL3D simulations of LHCD and demonstrated that full-wave effects, such as diffraction and interference, only weakly affect LH wave power deposition and current drive profiles. Diffractional broadening was quantified and found to be much smaller than the spectral broadening from finite launch spectrum width. In fact, diffraction was only noticable in situations where the wave was damped in 2-3 passes and in weaker damping the LH spectral gap closure was completely dominated by the geometric upshift effect. Thus, it was determined full-wave effects cannot account for the differences between simulations of LHCD by raytracing/Fokker-Planck models and experiments, and despite breakdowns in the raytracing approximation that can occur at cutoffs and caustics raytracing/FP simulations robustly reproduced full-wave/FP results. The primary discrepancy between raytracing and full-wave simulations, the reduced current drive efficiency observed in full-wave simulation, may be explained by systemic error introduced from the non-positive definiteness of the full-wave quasilinear diffusion coefficient. This presents an obvious topic for future work, but as positive definite formulations of the quasilinear diffusion coefficient are $> \mathcal{O}(10^6)$ more expensive than the standard formulation of $D_{ql}$ used here \citep{LeePPCF2017}, they were deemed too computationally demanding for implementation in the LHCD problem. Another complication of implementing such a form of the quasilinear diffusion is it may be necessary to perform a correction for the parallel magnetic field inhomogeneity to the $\textrm{Im}[\chi_{zz}]$ to ensure self-consistency (if the inhomogeneous correction is physically relevant outside of ensuring $D_{ql}$ positive-definiteness). However, if positive definite forms of $D_{ql}$ could be implemented they would have the added benefit of self-consistently including evanescent modes we have removed in the simulations here as they do not model resonant damping with a delta function. 

Our TORLH/CQL3D results are the first full-wave LHCD simulations to demonstrate both precise RF power deposition and integrated power agreement between the full-wave electric field and the Fokker-Planck simulation results without the use of numerical power rescaling factors. To do this many of the non-Maxwellian components in the TORLH/CQL3D simulations were rewritten with a focus on improving self-consistency. We implemented noise reduction and fixed errors in the quasilinear diffusion coefficient formulation, improved the $\textrm{Im}[\chi_{zz}]$ formulation in TORLH to use $f$ derivatives matched to those in CQL3D, implemented a robust iteration framework using the IPS, and improved TORLH's performance so that it could be run at higher resolution to suppress spectral pollution. This allowed us to perform the very large, 8000+ cores for 36+ hours, simulations here. In total, this study used $\sim 100$ million total CPU hours over 4 years on NERSC. 

In summary, full-wave effects' influence on LHCD power and current deposition profiles appears to be small. Our simulations indicate raytracing/FP simulations accurately reproduce LH wave propagation and damping in the core of tokamaks, but in some cases with stronger damping there may be a small correction related to diffraction that will usually be smaller than other corrections, like finite launcher spectrum width. If LHCD is moderately damped, and a launcher with a very small spectral width is used or very accurate simulations are required, an effective way to account for diffractional broadening might be the use of a higher-order-accurate WKB method such as beamtracing that captures diffraction \citep{Bertelli2012,Poli2018}.

\section*{Acknowledgements}
This work was supported by Scientific Discovery Through Advanced Computing (SCIDAC) Contract No. DE-SC0018090 and Department of Energy grant: DE-FG02-91ER54109. This research used resources of the National Energy Research Scientific Computing Center, a DOE Office of Science User Facility supported by the Office of Science of the U.S. Department of Energy under Contract No. DE-AC02-05CH11231 using NERSC award FES-ERCAP0020035.

\appendix

\section{$\textrm{Im}[\chi_{zz}]$ Formulation From CQL3D Plasma Distributions} \label{sec:chizzform}

To begin our derivation of a non-Maxwellian $\textrm{Im}[\chi_{zz,e}]$ appropriate for use in TORLH we take the $n=0$ $zz$ components of the relativistic hot plasma dielectric for a general distribution function from \citet{Stix}:
\begin{equation}
    \chi_{zz,e} = 2\pi \frac{\omega_{pe,0}^2}{\omega \Omega_{e,0}} \int_0^\infty dp_\perp p_\perp \int_{-\infty}^\infty p_\parallel dp_\parallel \left( \frac{\Omega_e}{\omega - k_\parallel v_\parallel} \right) J_0^2\left(k_\perp v_\perp / \Omega_e \right) \frac{\partial f_e}{\partial p_\parallel}
\end{equation}
Where relativistic momentum $\vec{p}=\gamma m_0 \vec{v}$ and we have denoted quantities which use the rest mass, $m_0$, with subscript $0$ and quantities that utilize the relativistic mass $\gamma m_0$ without a subscript, i.e. $\Omega = \Omega_o/\gamma$. Integrating in $p_\parallel$ and taking the imaginary component from the residue of the pole in the resonant denominator yields:
\begin{equation}\label{eq:imchipretabl}
    \textrm{Im}[\chi_{zz,e}] = -2\pi^2 m_0 \frac{\omega_{pe,0}^2}{k_\parallel^2}\int_0^\infty dp_\perp p_\perp J_0^2\left(\frac{k_\perp p_\perp}{m_0 \Omega_{e,0}}\right) \frac{\partial p_\parallel}{\partial v_\parallel} \left. \frac{\partial f_e}{\partial p_\parallel} \right|_{v_\parallel = \omega/k_\parallel}
\end{equation}
We want to create a lookup table in TORLH that is a function of, radial location $\psi$, flux surface angle $\theta$, and $N_\parallel$ using derivatives of the CQL3D distribution function at the outboard midplane $F_0(\psi,u,\vartheta)$. We transform (\ref{eq:imchipretabl}) into a more convenient form in terms of CQL3D \& TORLH variables with the following relations: 
\begin{subeqnarray}
    u_n &=& p/m_oc = \sqrt{u_{\parallel n}^2 + u_{\perp n}^2} = \sqrt{\frac{N_\parallel^2u_\perp^2 + 1}{N_\parallel^2 -1}}\\
    \gamma &=& \sqrt{1+u_n^2} \\
    u_{\parallel n} &=& \sigma_\pm \left( \frac{1+u_{\perp n}^2}{N_\parallel^2 - 1} \right)^{1/2} \\
    u_{\perp n} &=& p_\perp / m_o c \\
    \frac{\partial p_\parallel}{\partial v_\parallel} &=& \frac{m_0 \gamma^3}{1+u_\perp^2} 
\end{subeqnarray}
where $\vec{u}_n$ is the normalized momentum per rest mass, and $\sigma_\pm = \textrm{Sign}(\omega/k_\parallel)$. Using these equations we may rewrite (\ref{eq:imchipretabl}) in terms of our normalized table variables:
\begin{subeqnarray}
    \textrm{Im}[\chi_{zz}](\psi,\theta,n_\parallel) &=& -\alpha \int_0^\infty du_\perp^2 J_0^2\left(\frac{k_\perp u_\perp c}{\Omega_{e,0}} \right)(1+u^2_\perp)^{1/2}\frac{\partial f_e}{\partial p_\parallel} |_{v_\parallel = \omega/k_\parallel} \\
    \alpha &=& \pi^2 \frac{\omega_{pe,0}^2}{\omega^2}(m_{e,0} c)^4\frac{N_\parallel}{(N_\parallel^2-1)^{3/2}}
\end{subeqnarray}
Finally, we must rewrite the derivative of the distribution function $\partial f / \partial p_\parallel$ in terms of the distribution function at the outboard midplane that is output by CQL3D this requires that we relate $f(p_\parallel,p_\perp,\theta,\psi)$ to $F_0(u,\vartheta_0(\vartheta,\theta),\psi)$, where $\vartheta$ is the velocity-space pitch angle. These quantities may be related by applying conservation of momentum and magnetic moment yielding equation: 
\begin{equation}
    \frac{\partial f}{\partial p_\parallel} = \frac{\cos\vartheta}{m_{e,0}}\left( \frac{\partial F_0}{\partial u} - \frac{\tan \vartheta_0}{u} \frac{\partial F_0}{\partial \vartheta_0}\right)
\end{equation}
With this our derivation of $\chi_{zz}$ is complete. To construct the lookup table a grid in $\theta$ and $N_\parallel$ must be specified (the $\psi$ grid is set automatically based on the CQL3D flux surface locations), and an integration grid in $u_\perp$ must also be specified. $\chi_{zz}$ is then calculated from the CQL3D distribution function at each gridpoint. In order to find $\partial f/ \partial p_\parallel$ the nearest neighbor point on the CQL3D grid point corresponding to the Landau resonance location at a given $u_\perp$ and $u_\parallel$ is used (no noticeable improvement in self-consistency was observed with more complicated interpolations). 

\bibliographystyle{jpp}

\bibliography{main}

\end{document}